\documentclass[preprint,nofootinbib,amsmath,amssymb]{revtex4-1}

\usepackage{graphicx}
\usepackage{multirow}

\begin{document}

\title{Confronting Phantom Dark Energy with Observations}

\author{Pao-Yu Wang}
\affiliation{Department of Physics, National Taiwan University, Taipei 10617, Taiwan, R.O.C.}

\author{Chien-Wen Chen}

\affiliation{Leung Center for Cosmology and Particle Astrophysics
(LeCosPA), National Taiwan University, Taipei 10617, Taiwan, R.O.C.}

\author{Pisin Chen}
\email{chen@slac.stanford.edu} %
 \affiliation{LeCosPA, Department of Physics, and
Graduate Institute of Astrophysics,
National Taiwan University, Taipei 10617, Taiwan, R.O.C.} 
\affiliation{Kavli Institute for Particle Astrophysics and
Cosmology, SLAC National Accelerator Laboratory, Menlo Park, CA
94025, U.S.A.} %

\begin{abstract}

We confront two types of phantom dark energy potential with observational data. The models we consider are the power-law potential, $V\propto\phi^{\mu}$, and the exponential potential, $V\propto \exp\left(\lambda\phi/M_P\right)$. We fit the models to the latest observations from SN-Ia, CMB and BAO, and obtain tight constraints on parameter spaces. Furthermore, we apply the goodness-of-fit and the information criteria to compare the fitting results from phantom models with that from the cosmological constant and the quintessence models presented in our previous work. The results show that the cosmological constant is statistically most preferred, while the phantom dark energy fits slightly better than the quintessence does.

\end{abstract}


\maketitle

\section{INTORDUCTION\label{sec:introduction}}

Observations over the past dozen years have shown that the universe is currently under accelerating expansion (see~\cite{Frieman:2008sn},~\cite{Caldwell:2009ix} for reviews). Under the framework of general relativity and standard cosmology, a new form of exotic energy with negative pressure ($p<-\rho/3$) is required to explain this phenomenon. Current observations suggested this so called "dark energy" made up about 73\% of the energy density of the universe~\cite{Suzuki:2011hu}~\cite{Komatsu:2010fb}~\cite{Blake:2011en}. So far $w_{DE}$ has been constrained to be very close to $-1$ assuming it is constant and the universe is flat. This seems to suggest the observation data prefer a cosmological constant as dark energy. However, a parametrized dark energy $w=w_0+w_a(1-a)$ has $\sim100\%$ uncertainty in dynamical parameter $w_a$ , when compared to data. This means dynamical dark energy models are not excluded.

Several dark energy model have been proposed in order to explain the cosmic acceleration. In addition to the most discussed cosmological constant, dynamically evolving scalar-field dark energy has been widely studied (for examples, see~\cite{Ratra:1987rm}~\cite{Caldwell:1998ii}~\cite{Zlatev:1999tr}). Quintessence is a specific case of scalar-field dark energy with canonical kinetic terms, which admits $-1<w_{\phi}<1$. It has drawn much attention because it can in principle provide the "tracker" property -- a property that makes the energy density today insensitive to its initial condition, i.e., the initial energy density value $\rho_{\phi}$ ~\cite{Zlatev:1999tr},~\cite{Steinhardt:1999nw}. While fine-tuning of potential parameter is still required~\cite{Singh:2003vx}, tracker quintessence can alleviate the cosmic coincidence problem because a precise setting of initial energy density ratio for matter and dark energy is no longer required.

Present observational constraint on $w_{DE}$ still allows for $w_{DE}<-1$. For example, results from WiggleZ ~\cite{Blake:2011en} showed current constraints on constant $w$ is $-1.114<w<-0.954$ after combining with the latest SN-Ia, CMB and BAO data. We note that while quintessence only allows for$-1<w_{\phi}<1$, phantom scalar-field, another dynamical DE model, proposed by Caldwell, Carroll et al~\cite{Caldwell:1999ew}, that invokes a negative kinetic energy, can satisfy $w_{\phi}<-1$. While there exist several known theoretical difficulties for phantom scalar-field dark energy model such as the violation of null dominant energy condition (NDEC)~\cite{Carroll:2003st} and a possible Big Rip phase in the future~\cite{Caldwell:2003vq}~\cite{Scherrer:2004eq}, it is still very worth while to confront it directly and independently against the cosmological observations, especially since the current best-fit for equation-of-state is smaller than $-1$(see~\cite{Caldwell:2003vq}~\cite{Guo:2004ae}~\cite{Sun:2004bf} for examples).

In our previous work~\cite{Wang:2011bi}, we have tested several tracker quintessence models with observational data. The result showed that the best-fit of the inverse-power-law potential and the inverse-exponential potential models both reduced to the cosmological constant. Motivated by this and the implication of $w_{DE}<-1$ from observations, in this paper we put the phantom dark energy models to the test. We consider two specific scalar-field phantom potentials: the power-law potential~\cite{Guo:2004ae} and the exponential potential~\cite{Li:2003ft}, each with one free parameter. The reason to choose these potentials is that they also possess the attractor-like property: insensitive to initial conditions. We take the model-based approach to confront the models with observational data. The data we use includes the latest Type-Ia supernova (SN-Ia) compilation set, the cosmic microwave background (CMB) and the baryon acoustic oscillations (BAO) observations. We also confront these models with the cosmological constant and the quintessence scalar-field models by using the Goodness-of-Fit test and the information criteria to assess the merit of each model.

\section{TRACKER AND ATTRACTOR PHANTOM}
\subsection{Phantom Formalism}

Phantom dark energy with equation of state $w<-1$ is achieved by introducing a negative kinetic energy term in the action. In this way phantom scalar field is slowly "rolling up" the potential. The energy density and pressure of phantom field can then be given as
\begin{equation} \rho_{\phi} =
-\frac{1}{2}\dot{\phi}^2 + V(\phi) \label{three},\end{equation}
\begin{equation} p_{\phi} =
-\frac{1}{2}\dot{\phi}^2 - V(\phi)\label{four}.\end{equation}
with $w_{\phi}(z)=p_{\phi}/\rho_{\phi}<-1$ in the range $0<\phi^2/2<V(\phi)$. The evolution of the phantom field is governed by its equation of motion:
\begin{equation} 
\ddot{\phi}+ 3H\dot{\phi}-\frac{dV}{d\phi} = 0,
\label{EoM}
\end{equation}
in which $H$ denotes the Hubble expansion rate, the dot denotes the derivative w.r.t. the physical time. 
Assuming a flat universe, the Friedmann equation can be written as
\begin{eqnarray} H^2(z)&=&\frac{8\pi G}{3}\left(\rho_r(z)+\rho_m(z)+\rho_\phi(z)\right)\nonumber \\
&=&H_0^2 \left[ \Omega_{r} (1+z)^4+\Omega_{m} (1+z)^3 +  \Omega_{\phi} \exp
\left( 3 \int_{0}^{z} \left[ 1 + w_{\phi}(z') \right]
\frac{dz'}{1+z'} \right) \right],
\label{FDeq}\end{eqnarray}
where $\rho_r(z)$ is the radiation energy density, $\rho_m(z)$ is the matter energy density, $\rho_\phi(z)$ is the scalar field energy density, $H_0$ is the Hubble constant.

\subsection{Tracker Phantom}

In the quintessence scenario, there exists a special ``tracker solution" to which other solutions would converge ~\cite{Zlatev:1999tr}~\cite{Steinhardt:1999nw}. A wide range of initial conditions for $\phi$ and $\dot{\phi}$ will approach a common evolutionary track of $\rho_{\phi}$ and $w_{\phi}$; this means the tracker field model is insensitive to its initial conditions. A very large range of initial values of $\rho_{\phi}^{i}$ is thus allowed without changing cosmic history. This property makes the tracker model very interesting to study, because it can alleviate cosmic coincidence problem. Conditions for tracker quintessence are such that $\Gamma\equiv V''V/(V')^2>1$ and is nearly constant ($|\Gamma'/\Gamma(V'/V)|\ll 1$) for a wide range of plausible field initial conditions. Here the prime denotes the derivative w.r.t $\phi$. Under these conditions, the tracker solution for quintessence can be approximated as~\cite{Steinhardt:1999nw}
\begin{equation}
w_{\phi}\approx \frac{w_B -2(\Gamma-1)}{1+2(\Gamma-1)},
\label{trackersol}
\end{equation}
where $w_B$ is the background dominant component in the equation-of-state. The fact that $\Gamma>1$ ensures that $w_{\phi}<w_B$, so that at late times dark energy will eventually take over and become dominant. \\

\begin{figure}[h]
\includegraphics[width=1\linewidth]{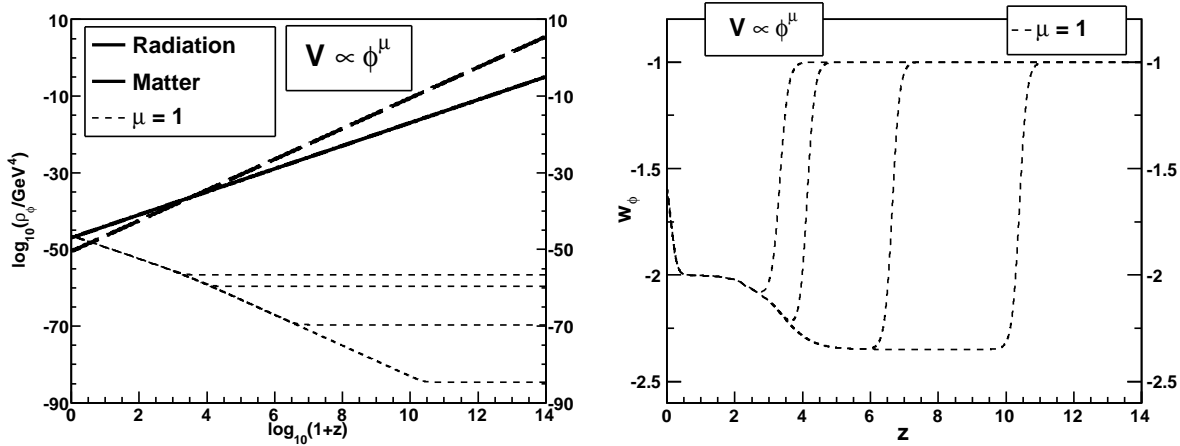}
\caption{Example of tracker phantom, with power-law potential.
\protect\label{combinePa}}
\end{figure}

The tracker solution for phantom has been studied in~\cite{Hao:2003th}~\cite{Chiba:2005tj}. Its tracker condition is closely related to that for quintessence: $\Gamma\equiv V''V/(V')^2<1/2$, and is nearly constant. Because $w_{\phi}\leq-1$ all the time, the energy density of phantom dark energy either stays the same or grows with time. This ensures that the dark energy density will eventually take over and become the dominant substance at late times. The form of its tracker solution is the same as in Eq.~(\ref{trackersol}). It is evident that the tracker phantom models are also insensitive to initial conditions for phantom scalar field. One example of tracker phantom is a power-law potential $V\propto \phi^{\mu}$ with $0<\mu<2$. In Fig.~\ref{combinePa}, we show an example for the tracking behavior of the power-law potential phantom.

\subsection{Attractor Phantom}

The tracker solution is just one type of ``attractor solutions" for scalar fields~\cite{Hao:2003th}~\cite{Ng:2001hs}. Attractor solutions are the stable critical points of autonomous equations (re-written from the equation of motion and Einstein equation) to which different initial conditions converge (see proof in ~\cite{Hao:2003th}~\cite{Ng:2001hs}~\cite{Hao:2003ww}). In contrast, the tracker solution is not a usual attractor because its critical point is not fixed, but changes with time when the background fluid dominates. Other than the tracker, there are two additional types of late time attractor solution for the phantom scalar field: the Big Rip attractor and the de Sitter attractor. These two attractors can only be reached in the future (when $\Omega_{\phi}=1$); however we found numerically that the exponential potential model in the Big Rip attractor case remains insensitive to some range of the initial values of $\phi$ and $\dot{\phi}$, as long as $\phi_{i}\ll M_{P}$. This allows certain range of initial $\rho_{\phi}$, and is thus still worth looking into.

The Big Rip attractor acquired its name because the attractor solution approaches $w_{\phi}<-1$ in the future. This will cause a catastrophic ``big rip", where the energy density and the scale factor will blow up within a finite time~\cite{Caldwell:2003vq}~\cite{Chimento:2004ps}. Although this may seem an unappealing feature, it is theoretically permissible and is therefore worthy of investigation. The condition for the Big Rip attractor is $\Gamma\simeq 1$ and $\lambda\equiv -V'/\kappa V\neq 0$~\cite{Hao:2003th}, $\kappa^2=8\pi G$. One example is the exponential potential $V\propto \exp\left(\phi/M_P\right)$ mentioned above.

In this paper, we consider two phantom scalar-field dark energy models: the power-law potential $V\propto\phi^{\mu}$ and the exponential potential $V\propto \exp\left(\lambda\phi/M_P\right)$. For the power-law potential phantom, it has the tracker property. For the exponential potential phantom, there exists an attractor solution that approaches the big rip in the future. We analyze these two models with observational data, obtain constraints on the model parameter, and assess the merit of the models derived from the best-fit results.

\section{Data}

We use observational data from SN-Ia, CMB and BAO described below.

\subsection{Type-Ia Supernovae}

We use the latest SNIa dataset, Union2.1 compilations, released by the Supernova Cosmology Project that contains 580 SN-Ia in the the redshift range of $0.02<z<1.5$~\cite{Suzuki:2011hu}. This compilation includes supernova data from~\cite{Hamuy:1996su}\textrm{--}\cite{Kessler:2009ys}. The dataset provides the distance modulus that contains information of the luminosity distance that can be used to constrain the dark energy.

The distance modulus is defined as following:
\begin{equation}
\mu_{th}(z)=5\log_{10}\left(\frac{d_L(z)}{Mpc}\right)+25=5\log_{10}\left(D_L(z)\right)+\mu_0,
\end{equation}
where $D_L(z)=H_0 d_L(z)$ is the Hubble-free luminosity distance. We marginalize $\chi^2_{SNIa}$ over the nuisance parameter $\mu_0$ by minimizing it with respect to $\mu_0$. The marginalized $\chi^2_{SNIa}$ is~\cite{DiPietro:2002cz}\textrm{--}\cite{Wei:2009ry}
\begin{equation}
\tilde{\chi}^2_{SNIa}=A-\frac{B^2}{C},
\end{equation}
where
\begin{eqnarray}
A&=&\sum_{ij}\Big[5\log_{10}D_L(z_i,par)-\mu_{obs}(z_i)\Big]C^{-1}_{ij}\Big[5\log_{10}D_L(z_j,par)-\mu_{obs}(z_j)\Big],\\
B&=&\sum_{ij}\Big[5\log_{10}D_L(z_i,par)-\mu_{obs}(z_i)\Big]C^{-1}_{ij},\\
C&=&\sum_{ij}C^{-1}_{ij}.
\end{eqnarray}

\subsection{Cosmic Microwave Background}

The 7-year WMAP data provides the ``distance prior" that can be used to constrain dark energy~\cite{Komatsu:2010fb}. The distance prior includes the CMB shift parameter $R=1.725\pm0.018$ given by
\begin{equation}
R=\sqrt{\Omega_m H_0^2}\left(1+z_*\right)D_A\left(z_*\right),
\end{equation}
and ``acoustic scale" $l_A=302.09\pm0.76$ given by
\begin{equation}
l_A= \left(1+z_*\right)\frac{\pi D_A\left(z_*\right)}{r_s\left(z_*\right)},
\end{equation}
where $z_*$ is the redshift at decoupling, $D_A$ is the physical angular diameter distance, and $r_s$ is the comoving sound horizon. We use the fitting formula proposed by Hu and Sugiyama~\cite{Hu:1995en}:
\begin{equation}
z_*=1048\left[1+0.00124(\Omega_b h^2)^{-0.738}\right]\left[1+g_1(\Omega_m h^2)^{g_2}\right],
\end{equation}
\begin{eqnarray}
g_1=\frac{0.0783(\Omega_b h^2)^{-0.238}}{1+39.5(\Omega_b h^2)^{0.763}},
\end{eqnarray}
\begin{eqnarray}
g_2=\frac{0.560}{1+21.1(\Omega_b h^2)^{1.81}}.
\end{eqnarray}
The comoving sound horizon is
\begin{equation}
r_s(z)=\frac{1}{\sqrt{3}}\int^{1/(1+z)}_{0}\frac{da}{a^2H(a)\sqrt{1+(3\Omega_b/4\Omega_{\gamma})a}},
\end{equation}
where $\Omega_b$ is the baryon density and $\Omega_{\gamma}$ is the photon density. We construct $\chi^2_{CMB}=\sum_{ij}[x_i-x^{Obs}_i]C^{-1}_{ij}[x_j-x^{Obs}_j]$, where $C^{-1}_{ij}$ is the inverse covariance matrix given in~\cite{Komatsu:2010fb}, and $x_i=(l_A, R, z_*)$.

\subsection{Baryon Acoustic Oscillations}
We followed~\cite{Blake:2011en} and use three sets of BAO distance dataset: 6dFGS, SDSS and WiggleZ, for our study.

We use the joint analysis of the Two Degree Field Galaxy Redshift Survey (2dFGRS) data~\cite{Cole:2005sx} and the Sloan Digital Sky Survey (SDSS) Data Release 7, which provides two distance measures of $d_{0.35}=r_s(z_d)/D_V(0.35)=0.1097\pm0.0036$ and $d_{0.2}=r_s(z_d)/D_V(0.2)=0.1905\pm0.0061$~\cite{Percival:2009xn}, where $r_s(z_d)$ is the acoustic sound horizon at the drag epoch, $D_V= \left[(1+z)^2D^2_A(z)/H(z)\right]^{1/3}$. The fitting formula for $z_d$ is defined by Eisenstein \& Hu~\cite{Eisenstein:1997ik}. The $\chi^2_{BAO_1}$ is $\sum_{ij}[d_i-d^{obs}_i]C^{-1}_{ij}[d_j-d^{obs}_j]$, where $d_i=(d_{0.2}, d_{0.35})$ and
\begin{equation}
C^{-1}=
\left(
\begin{array}{cc}
30124 & -17227 \\
-17227 & 86977 \end{array}
\right).
\end{equation}
The fitting formula for $z_d$ has the form:
\begin{equation}
z_d=\frac{1291(\Omega_mh^2)^{0.251}}{1+0.659(\Omega_mh^2)^{0.828}}\left[1+b_1(\Omega_bh^2)^{b_2}\right],
\end{equation}
where
\begin{eqnarray}
b_1=0.313(\Omega_mh^2)^{-0.419}\left[1+0.607(\Omega_mh^2)^{0.674}\right], \quad\quad
b_2=0.238(\Omega_m h^2)^{0.223}.
\end{eqnarray}
The second is the 6dFGS data in~\cite{Beutler:2011hx}, which provides $d_z(0.106)=0.336\pm0.015$. We thus have $\chi^2_{BAO_2}=[d_z(0.106,par)-0.336]^2/(0.015)^2$.

Finally, we include the result from the WiggleZ Survey~\cite{Blake:2011en}. WiggleZ provides three correlated measurements: $\vec{A}^{obs}=\left(A(z=0.44), A(z=0.6), A(z=0.73)\right)=(0.474, 0.442, 0.424)$, with the inverse covariant matrix  
\begin{equation}
C^{-1}=\left(
\begin{array}{ccc}
1040.3 & -807.5 & 336.8 \\
& 3720.3 & -1551.9 \\
& & 2914.9
\end{array}
\right),
\end{equation}
and $A(z)$ defined as
\begin{equation}
A(z)\equiv \frac{100D_V(z)\sqrt{\Omega_m h^2}}{cz}.
\end{equation}
The $\chi^2_{BAO_3}$ can be written as $[A^{obs}_{i}-A_{i}]C^{-1}_{ij}[A^{obs}_{j}-A_{j}]$. It is obvious that $\chi^2_{BAO}= \chi^2_{BAO_1}+\chi^2_{BAO_2}+\chi^2_{BAO_3}$.

\subsection{Prior}
 For the radiation, we fix $\Omega_{\gamma}=2.469\times10^{-5}/h^2$, and use the relation $\Omega_r=\Omega_{\gamma}(1+0.2271N_{eff})$ as the radiation energy density~\cite{Komatsu:2008hk}. $N_{eff}$ is the effective number of neutrino species and is taken to be $3.04$~\cite{Komatsu:2008hk}. We further impose the prior of $H_0=73.8\pm2.4$~km s$^{-1}$ Mpc$^{-1}$ from~\cite{Riess:2011yx}. This prior is an independent and complementary constraint on parameter $h$. The total chi-square $\chi^2_{total}=\tilde\chi^2_{SNIa}+\chi^2_{CMB}+\chi^2_{BAO} + \chi^2_{H_0}$ is marginalized over $\Omega_bh^2$ and the reduced Hubble constant $h$ by minimizing $\chi^2_{total}$ with respect to $\Omega_bh^2$ and $h$~\cite{Komatsu:2008hk}.

\section{Observational Constraints on Tracker and Attractor Phantoms}
We consider two potential forms, the power-law potential $V=V_1\left(\phi/M_P\right)^{\mu}$ and the exponential potential $V=V_2 \exp\left(\lambda\phi/M_P\right)$. As mentioned above, the power-law potential corresponds to the tracker phantom, whereas the exponential potential admits a late time big rip attractor solution. Here $\mu$ and $\lambda$ are positive, dimensionless constants. The constant $V_1$ and $V_2$ are determined by requiring total energy density today equals to the critical energy density $\rho_c$ in a flat universe (i.e $\Omega_{tot}=1$). We calculate our $\chi^2$ by solving Eq.~(\ref{EoM}) and Friedmann equation Eq.~(\ref{FDeq}) numerically.

\begin{figure}[h]
\includegraphics[width=1\linewidth]{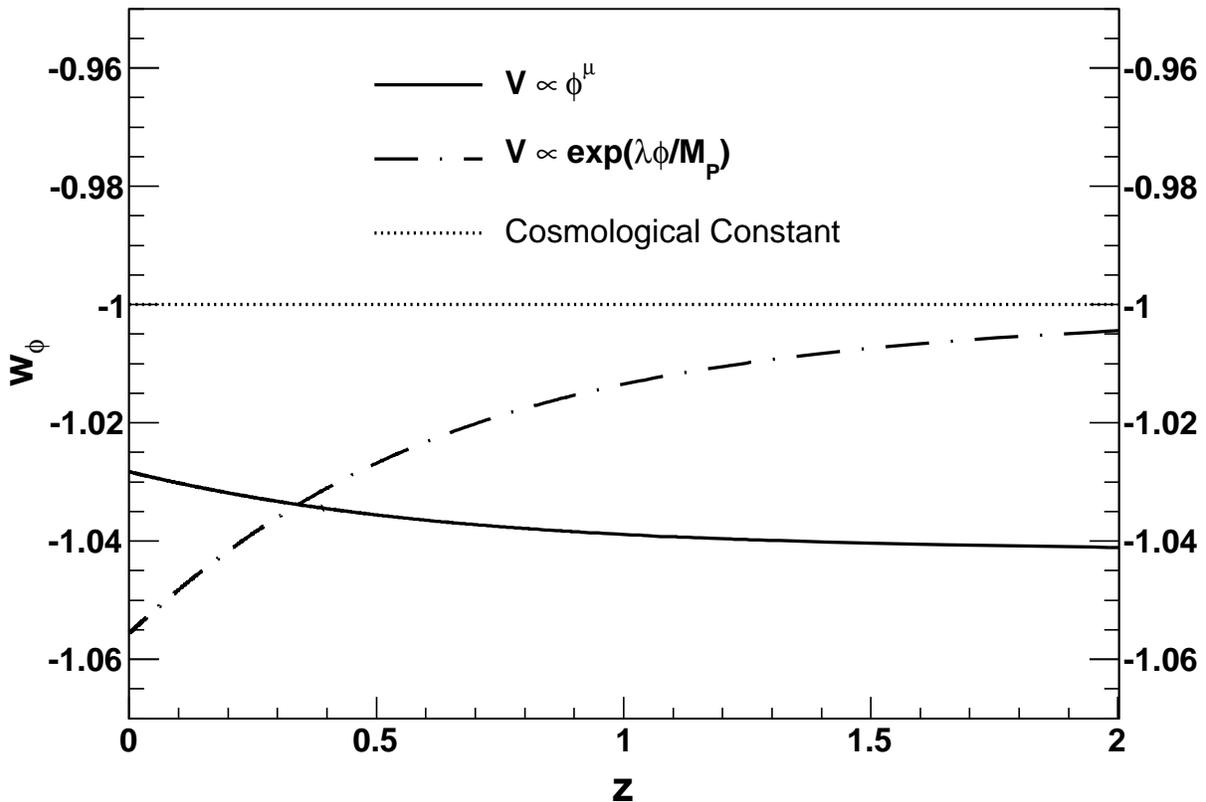}
\caption{The late time evolution of $w_{\phi}(z)$ corresponding to the best-fit parameters. Solid curve corresponds to $V\propto \phi^{\mu}$, dot-dash curve corresponds to $V\propto \exp\left(\lambda\phi/M_P\right)$.
\protect\label{wbest}}
\end{figure}

\begin{figure}[ht]
\includegraphics[width=1\linewidth]{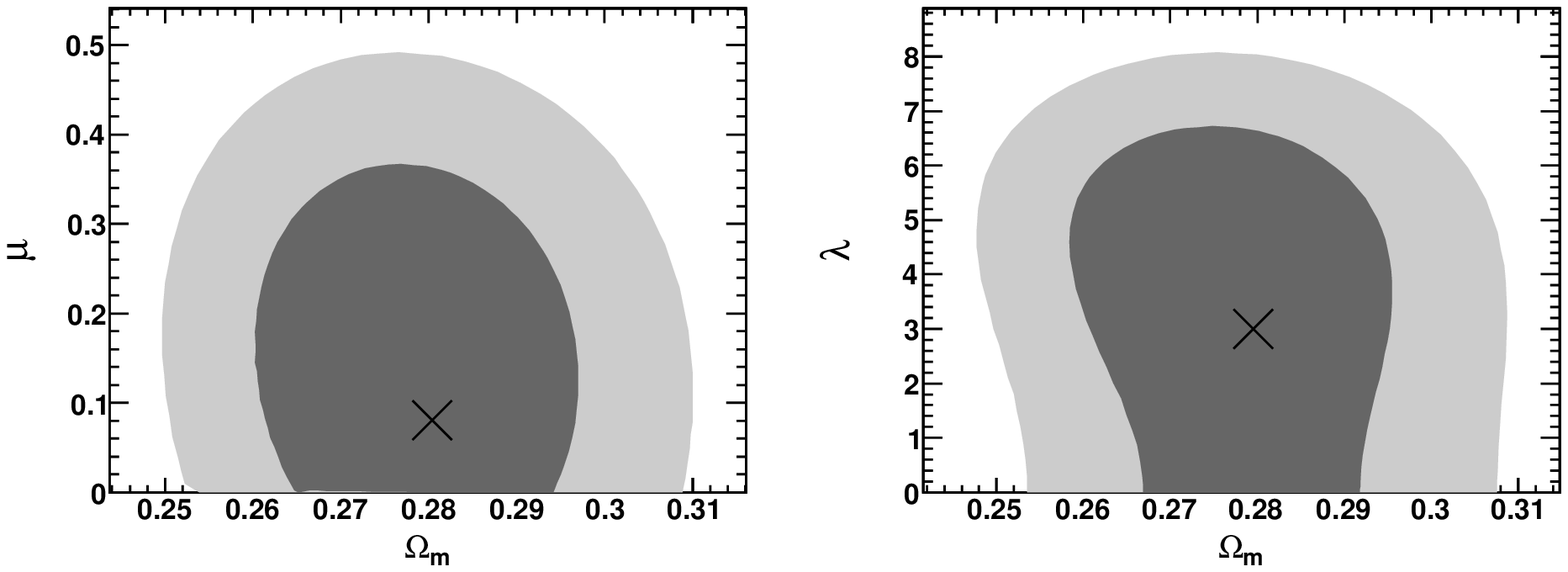}
\caption{Joint constraints on ($\Omega_m, \mu$) and ($\Omega_m, \lambda$). The dark gray and the light gray regions correspond to the $68.3\%$ and $95.4\%$ confidence regions, respectively.
\protect\label{joint}}
\end{figure}

The results are given in Table~\ref{table:results}. For comparison, we also provide results of two quintessence models, the inverse-power-law potential and the inverse-exponential potential. The best-fit equation of states for two phantom models at late times are given in Fig.~\ref{wbest}, and the two-parameter likelihood contours ($(\Omega_m, \alpha),(\Omega_m,\lambda)$) are given in Fig.~\ref{joint}. 

In order to test the merit of the model, we perform the goodness-of-fit (GoF) test to all models. The meaning for GoF is, assuming a model to be true, the probability of finding a new set of data that gives worse $\chi^2$ than that deduced by the current data. The higher the GoF is, the more viable is the model. Explicitly, it is defined as $\Gamma(\nu/2,\chi^2/2)/\Gamma(\nu/2)$, where $\Gamma(\nu/2,\chi^2/2)$ is the upper incomplete gamma function and $\nu$ is the degrees of freedom. 

To further assess the relative merit between models, we invoke the ``information criteria". The information criteria (IC) is a set of statistical considerations that take both data fitting and model complexity into account; they favor models with better fit and fewer parameters. The application of the information criteria to cosmology was first launched by Liddle~\cite{Liddle:2004nh}. Here we consider two kinds of information criteria: the Bayesian information criterion (BIC)~\cite{Schwarz} and the Akaike information criterion (AIC)~\cite{Akaike}. The BIC is given by $\textrm{BIC}=-2\ln \mathcal{L}_{max} + k \ln N$, where $\mathcal{L}_{max}$ is the maximum likelihood, which is equivalent to the minimum $\chi^2$ for gaussian errors, $k$ is the number of parameters, and $N$ is the number of data points used in the fit. The second term $k\ln N$ serves as the ``penalty" to the model that invokes extra parameters. The AIC is defined as $\textrm{AIC}=-2\ln \mathcal{L}_{max} + 2k$. Both BIC and AIC favor smaller values while BIC charges stiffer penalty for extra parameters when $\ln N > 2$. Taking the BIC and AIC for the cosmological constant as the reference value, we compute the differences in BIC ($\Delta$BIC) and AIC ($\Delta$AIC) of other dark energy models. They are listed in Table~\ref{table:results}.

\begin{table}[ht]
\caption{Fitting Results}
\centering
\begin{tabular}{c|c|c|c c c}
\hline\hline
Model & Best-fit $\chi^2$ & Best-fit parameters~\footnote{the best-fit and the 68.3$\%$ confidence interval for each parameter ($\Delta \chi^2=1$) } & GoF & $\Delta$BIC & $\Delta$AIC \\
\hline
\multirow{4}{*}{Cosmological Constant} & $\chi^2_{tot}=550.41$ & \multirow{4}{*}{$\Omega_m=0.281^{+0.013}_{-0.012}$} &\multirow{4}{*}{$85.8\%$} & \multirow{4}{*}{0.0} & \multirow{4}{*}{0.0} \\ 
& $\chi^2_{SNIa}=545.24$ & & & & \\
& $\chi^2_{CMB}=0.19$ & & & & \\
& $\chi^2_{BAO}=2.32$ & & & & \\

\hline
& $\chi^2_{tot}=550.09$ & &\multirow{4}{*}{$85.4\%$} & \multirow{4}{*}{6.1} & \multirow{4}{*}{1.6} \\ 
Phantom with & $\chi^2_{SNIa}=545.66$ & $\Omega_m=0.280^{+0.013}_{-0.012}$ & & & \\
$V\propto \exp\left(\lambda\phi/M_P\right)$ & $\chi^2_{CMB}=0.37$ & $\lambda=2.99^{+2.10}$ & & & \\
& $\chi^2_{BAO}=2.06$ & & & & \\
\hline

& $\chi^2_{tot}=550.17$ & &\multirow{4}{*}{$85.3\%$} & \multirow{4}{*}{6.1} & \multirow{4}{*}{1.8} \\ 
Phantom with& $\chi^2_{SNIa}=545.53$ & $\Omega_m=0.280^{+0.013}_{-0.012}$ & & & \\
$V\propto \phi^{\mu}$ & $\chi^2_{CMB}=0.41$ & $\mu=0.081^{+0.154}$ & & & \\
& $\chi^2_{BAO}=2.17$ & & & & \\
\hline\hline

\hline
quintessence with& \multirow{2}{*}{ Same as cosmological constant} & $\Omega_m=0.281^{+0.013}_{-0.012}$ &\multirow{2}{*}{$85.1\%$} & \multirow{2}{*}{6.4} & \multirow{2}{*}{2.0} \\ 
$V\propto\phi^{-\alpha}$ &  & $\alpha=0^{+0.11}$ & & & \\

\hline
quintessence with & \multirow{2}{*}{ Same as cosmological constant} & $\Omega_m=0.281^{+0.013}_{-0.012}$ &\multirow{2}{*}{$85.1\%$} & \multirow{2}{*}{6.4} & \multirow{2}{*}{2.0} \\ 
$V\propto \exp\left(\beta M_P/\phi\right)$ &  & $\beta=0^{+0.010}$ & & &\\

\hline\hline

\end{tabular}
\label{table:results}
\end{table}

\section{DISCUSSION}

We have tested two potential forms of phantom scalar-field dark energy, the power-law $V=V_1\left(\phi/M_P\right)^{\mu}$ and the exponential potential, $V=V_2 \exp\left(\lambda\phi/M_P\right)$, with current observations. Tight model parameter constraints are obtained in Table~\ref{table:results}. We also provided results of cosmological constant and two types of quintessence potentials, the inverse-power law potential and the inverse-exponential potential. Cosmological constant yields the best goodness-of-fit and smallest information criteria. This shows that the cosmological constant is still the most preferred dark energy model among all that we have considered, even with the constraint $w>-1$ removed. Phantom models fit worse to observations than that with the cosmological constant, but are slightly better than the two quintessence potential models. Although the current observations still prefer the cosmological constant as the dark energy, phantom and quintessence models under considerations are only slightly worse in terms of GoF and the information criteria; all our models in Table~\ref{table:results} have GoF$\sim85\%$, $\Delta$BIC$\sim6$, and $\Delta$AIC$\sim2$. 

Another interesting result is the best-fit for phantoms. Unlike quintessence, which have the best-fit equivalent to $\Lambda$CDM~\cite{Wang:2011bi}, that for phantom models moves away from $w_{\phi}=-1$ (see Fig.~\ref{wbest}), as the best-fit for parameters $\lambda$ and $\mu$ are no longer zero (Fig.~\ref{joint}). For $V=V_1\left(\phi/M_P\right)^{\mu}$, the best-fit $w_{\phi}(0)\sim -1.03$; as for $V=V_2 \exp\left(\lambda\phi/M_P\right)$, we find $w_{\phi}(0)\sim -1.06$ for the best fit. This indicates that the current observations may prefer $w_{DE}<-1$.

In summary, the model-based approach we used in this paper suggests that the cosmological constant is more preferred, and dark energy models with $w<-1$ is preferred over $w>-1$. Notice that in both phantom models the cosmological constant cases are still inside $1-\sigma$ range. This means currently we cannot distinguish small-dynamical dark energy from the cosmological constant. Future observations from next generation dark energy probes are expected to constrain $w$ about ten times better than the present value~\cite{astro-ph/0609591}. More stringent constraints on the parameter space are thus expected to be obtained. By then we should be able to attain more insights into the physics of dark energy models with this model-based approach, or even rule out some of the models at a sufficient confidence level (see results with projected data in~\cite{Yashar:2008ju}~\cite{Barnard:2007ta}~\cite{Bozek:2007ti}~\cite{Abrahamse:2007te}).

%

%

\begin{acknowledgments}
This research is supported by the Taiwan National Science Council (NSC) under Project No. NSC98-2811- M-002-501, No. NSC98-2119-M-002-001, and the US Department of Energy under Contract No. DE-AC03- 76SF00515. We would also like to thank the NTU Leung Center for Cosmology and Particle Astrophysics for its support.
\end{acknowledgments}


\end{document}